\documentclass[a4paper]{jpconf}
\usepackage[pdftex]{graphicx}
\usepackage{caption}
\usepackage{subfigure}

\usepackage{amsthm}
\usepackage{amsmath}
\usepackage{amsfonts}

\theoremstyle{definition}

\newtheorem{theorem}{Theorem}[section]

\begin{document}
\title{Comparing the Behaviour of Deterministic and Stochastic Model of SIS Epidemic}

\author{Kurnia Susvitasari}

\address{School of Math. Science, University of Nottingham, The University Park, NG7 2RD, UK}

\ead{susvitasari@icloud.com}

\begin{abstract}
Studies about epidemic modelling have been conducted since before 19th century.
	Both deterministic and stochastiic model were used to capture the dynamic of 
	infection in the population.  The purpose of this project is to investigate the behaviour of the models when we set the basic reproduction number, $R_0$. This quantity is defined as the expected number of contacts made by a typical infective to susceptibles in the population. According to the epidemic threshold theory, when $R_0 \leq 1$, minor epidemic occurs with probability one in both approaches, but when $R_0 > 1$, the deterministic and stochastic models have different interpretation. In the deterministic approach, major epidemic occurs with probability one when $R_0 > 1$ and predicts that the disease will settle down to an endemic equilibrium. Stochastic models, on the other hand, identify that the minor epidemic can possibly occur. If it does, then the epidemic will die out quickly. Moreover, if we let the population size be large and the major epidemic occurs, then it will take off and then reach the endemic level and move randomly around the deterministic's equilibrium.
\end{abstract}

\section{Introduction} \label{sec: intro}
	Studies about epidemic modelling have been conducted since before 19th century.
	Both deterministic and stochastiic model were used to capture the dynamic of 
	infection in the population. Upon the study, some other parameters were introduced
	to generalised the model, for example the infection and recovery rates, as well
	as if the infection occurs in close or open population.
	
	One of the most important development in this study is threshold theorem. It states 
	that an epidemic can only occur if the initial number of susceptibles is larger than 
	some critical value which depends on the parameters of the model under 
	consideration. The threshold behaviour is usually expressed in terms of epidemic 
	basic reproduction number, $R_0$. This quantity is usually defined as the expected
	number of contacts made by a typical infective to susceptibles in the population. 
	It is important to note that in general epidemic modelling, both stochastic and
    deterministic models have similar threshold value, which is attained at $R_0 = 1$.
	
	According to N\aa sell \cite{nasell}, the proportion of infectives in the deterministic 
	model would converge to some equilibrium point as time approached infinity. There
	are only two stages of equilibrium, extinction or endemic. When $R_0$ below the 
	threshold point, both deterministic and stochastic approaches have converged
	to extinction. Unfortunately, different outcome happens when $R_0$ exceed 1. 
	
	\textbf{Motivation} The main purpose of this paper is to investigate the behaviour 
	of the SIS model when we set $R_0$. The model is named after the transition
	process, where each individual in the population is classified into some stages, i.e.
	\textit{susceptible (S)} and \textit{infective (I)}. The characteristics of the model
	are also studied, especially the conditions when the models enter the endemic-
	equilibrium level. 
	
\section{The SIS Epidemic Model}
	 
	 Suppose that we have a closed mixing population with initially $n$ individuals and $m$ 
	 infectives, where $m << n$. We define the SIS 
	 model as follows. 

\begin{figure}[ht!]
		\begin{center}
			\includegraphics[width=0.7\textwidth]{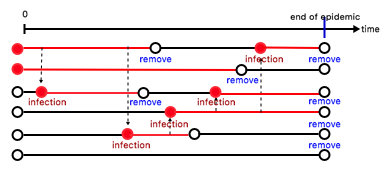}
		\end{center}
		\caption{SIS epidemic illustration} \label{fig:sis}
	\end{figure}

	At $t=0$, there are $n-m$ susceptibles who are vulnerable to the diseases. Each 
	infective contacts each susceptible according to independent Poisson processes with 
	rate $\lambda /n$  and if a susceptible is contacted, it immediately becomes an infective. 
	The infectives also have an infectious period which follow iid exponential distributions 
	with rate $\gamma$, and then removed, but once the infective’s infectious period is over, 
	the individual will immediately become vulnerable to the disease. Finally, the epidemic 
	ends when no infective is left in the population. The transition scheme of this model
	is illustrated in figure \ref{fig:sis}.
	
\subsection{The Transition Schemes of SIS Model}

	In this paper, we will assume that the population is closed. According to the figure
	\ref{fig:sis}, the SIS model consists of two processes, i.e. \textit{infection} and
	\textit{removal}.
	
	For $t \geq 0$, let $Y_n(t)$ be the number of infectives at time $t$. Suppose that 
	$Y_n(t) = i$, where $i \in \lbrace 1,2, \ldots n-1 \rbrace$. Then, there are $(n-i)i$
	contacts between infectivees and susceptibles in the population. Therefore, the
	probability of infection and removal occurence in interval $(t, t+dt)$ are respectively	
		\begin{align}
			\mathbb{P} \lbrace Y_n(t+dt) = i+1 | Y_n(t) = i \rbrace
				&=	\lambda/n \cdot (n-i) i \cdot dt + o(dt), 
				\label{eq: infection} \\
			\mathbb{P} \lbrace Y_n(t+dt) = i-1 | Y_n(t) = i \rbrace
				&=	\gamma i \cdot dt + o(dt).
				\label{eq: removal}
		\end{align}
		
\subsection{The Deterministic Approach of SIS Model}

	To capture the behaviour of the SIS model, suppose that we allow $Y_n(t)$ be
	non--integer. Considering there are two transisitions in the model, we can interpret
	the transition scheme as follows
		\begin{equation}
			Y_n(t+  dt) = 
				Y_n(t) + \left( \frac{\displaystyle \lambda}{\displaystyle n} Y_n(t) 
				\left[ n-Y_n(t) \right] - \gamma Y_n(t) \right) dt + o(dt) 
				\label{eq:sis_deter}
		\end{equation}
	which yields as $dt \rightarrow 0$,
		\begin{equation}
			\frac{\displaystyle dY_n(t)}{\displaystyle dt} = \frac{\displaystyle \lambda}
			{\displaystyle n} Y_n(t) \left[ n-Y_n(t) \right] - \gamma Y_n(t).
		\end{equation}

	Let $y(t) = \frac{\displaystyle Y_n(t)}{\displaystyle n}$ represent the proportion of 
	infective in the population at time $t$, then we can scale equation 
	(\ref{eq:sis_deter}) as
		\begin{align}
			\frac{dy(t)}{dt} &= \lambda y(t) \cdot (1-y(t)) - \gamma y(t), 		
			\label{eq:sis_deter_rate}
		\end{align}
	where $\lambda y(t) \cdot (1-y(t))$ and $\gamma y(t)$ are infection and removal
	rate transitions of $y(t)$ with solution
		$$y(t)=
			\begin{cases}
				\frac{\displaystyle (\lambda - \gamma) y(0) \cdot e^{(\lambda - \gamma)t}}	
				{\displaystyle \lambda - \gamma + \lambda y(0) \left( e^{(\lambda - \gamma)t} - 1 
				\right)} & \text{if } \lambda \neq \gamma, \\
				\frac{\displaystyle y(0)}{\displaystyle 1+ \lambda y(0) t} & \text{if } 
				\lambda = \gamma.
			\end{cases} $$
	
	Figure \ref{fig:sis.(n,m)=(1e3,10)} showed the behaviour of the processes with initial
	$(n,m)=(100,20)$. The stochastic processes behaved similar to deterministic 
	models, but in other figures, some stochastic processes were not mimicing the
	behaviour of deterministic model, especially when $R_0 > 1$. 
	
	From the simulations, there are some tentative conditions that can distinguish the
	behaviour of deterministic and stochastic approaches to model SIS. These conditions
	are determined by the value of $R_0$, as the threshold theorem stated:
		\begin{enumerate}
			\item both deterministic and stochastic approaches of SIS will go exinct
					if and only if $R_0 \leq 1$,
			\item in deterministic approach, as $t \rightarrow \infty$, the SIS model
					showed an outbreak of disease as $R_0 > 1$,
			\item in stochastic approach, there are two possibility of outcomes when
					$R_0 > 1$, i.e. the outbreak as deterministic approach or an extinction.
		\end{enumerate}
	
	Therefore, we will prove mathematically the three points above to determine the
	behaviour of SIS model based on its reproduction number.	
	
		\begin{figure}
			\centering
			\begin{minipage}{.5\textwidth}
				  \centering
				   \includegraphics[width=.75\textwidth, height=.65\textwidth]
				   	{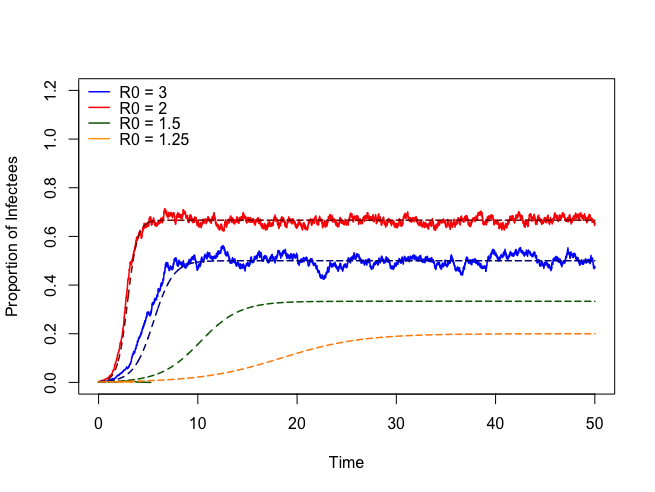}
 				 \caption{Initial susceptibles $=998$}
 				 \label{fig:sis.(n,m)=(1e3,2)}
			\end{minipage}%
			\begin{minipage}{.5\textwidth}
				  \centering
				   \includegraphics[width=.75\textwidth, height=.65\textwidth]
				   	{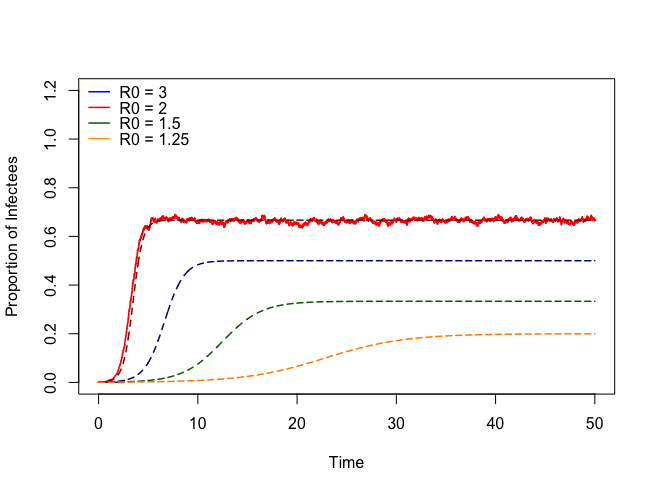}
				  \caption{Initial susceptibles $=2998$}
				  \label{fig:sis.(n,m)=(3e3,2)}
			\end{minipage} \\
				\centering
			\begin{minipage}{.5\textwidth}
				  \centering
				   \includegraphics[width=.75\textwidth, height=.65\textwidth]
				   {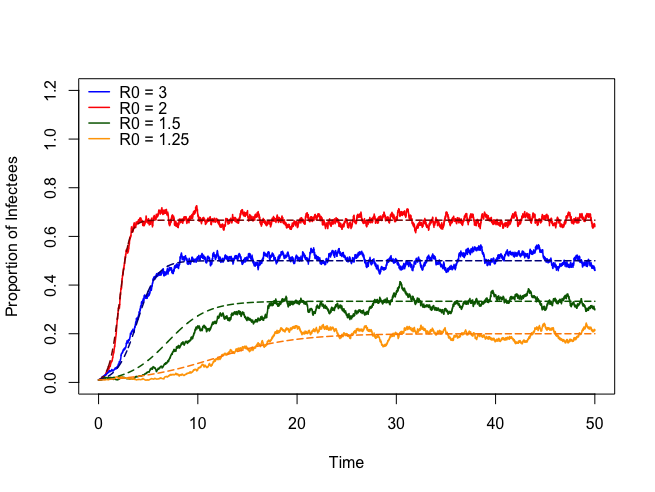}
				  \caption{$(S_0,I_0)=(990,10)$}
				  \label{fig:sis.(n,m)=(1e3,10)}
			\end{minipage}%
			\begin{minipage}{.5\textwidth}
				  \centering
				   \includegraphics[width=.75\textwidth, height=.65\textwidth]
				   	{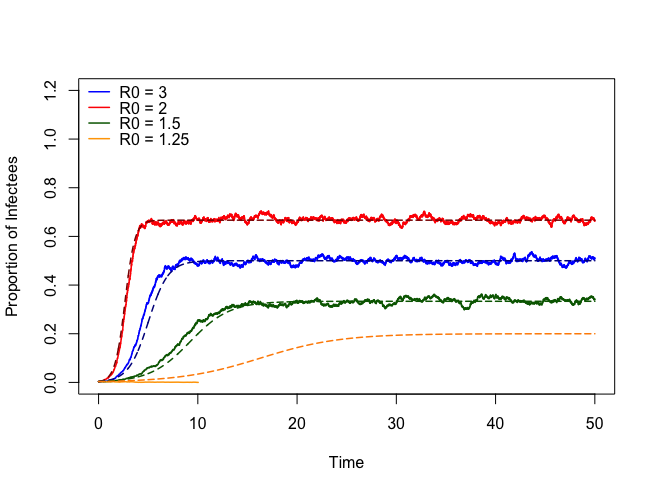}
				  \caption{ $(S_0,I_0)=(2990,10)$}
				  \label{fig:sis.(n,m)=(3e3,10)}
			\end{minipage}
			\end{figure}

\section{The Behaviour of SIS Model According to Deterministic Approach}

	 Recall equation (\ref{eq:sis_deter_rate}). Obviously, the major epidemic occurs if 
	 and only if $\frac{\displaystyle dy}{\displaystyle dt}>0$ and otherwise. At some time 
	 $t,$ the increment of $y(t)$ will reach stationary and the process will hit the 
	 equilibrium point as shown in Figure \ref{fig:sis.(n,m)=(1e3,2)} to 
	 \ref{fig:sis.(n,m)=(3e3,10)}. The roots of $\frac{\displaystyle dy}{\displaystyle dt} 
	 = 0,$ are $y(t) = 0$ when $R_0 = \lambda / \gamma \leq 1$ and $y(t) = 1-R_0^{-1}$
	 when $R_0$ exceeds 1. Therefore, according to the deterministic approach, the
     outbreak of epidemic happens when  $R_0$ exceeds 1 with probability one, and 
     otherwise.

\section{The Behaviour of SIS Model According to Stochastic Approach}

	It is easier to study stochastic model of SIS using simulation. In deterministic 
	approach, minor epidemic occurs if and only if $R_0 \leq 1$, whilst major epidemic 
	occurs 
	otherwise. This is because when we let $\frac{\displaystyle dy}{\displaystyle dt} 
	\leq 0 \left( \text{or } \frac{\displaystyle dy}{\displaystyle dt} > 0 \right)$, the 
	proportion of infection in (\ref{eq:sis_deter_rate}) decreases (increases) in 
	population upon time $t$ and then, the process will reach equilibrium.
	
	As $t \rightarrow \infty$, the stochastic model of SIS also behaved in similar way.
	In fact, some processes will mimic the deterministic model, but others are not.
	In the next section, we will mathematically show why and what conditions 
	stochastic model mimicing deterministic model.
	
\subsection{Density Dependence Population Process}

	Suppose that $\lbrace \boldsymbol{X}_n(t): t \geq 0 \rbrace$  is the CTMC process
	defined on $d$-dimensional integer lattice $\mathbb{Z}^d$  with finite number of possible
	transitions,  given by $q_{i,i+l}=n \beta_l \left( \frac{\displaystyle k}{\displaystyle n} 
	\right) $ where we define 
		$$\beta_l: \mathbb{Z}^d \rightarrow \left[ 0,\infty \right), l \in \Delta 
	\subseteq \mathbb{Z}^d, \sum_l \beta_l \left( \boldsymbol{x} \right) < \infty$$ 
	for all 
	$\boldsymbol{x} \in \mathbb{Z}^d$.
	
	According to Ethier and Kurtz \cite{ethier-kurtz}, if the the size of a certain population 
	was $k$, 
	then the density of that population was $\frac{\displaystyle k}{\displaystyle n}$ with 
	intensities approximated proportionally to its size. For example, 
	suppose that there are $i$-susceptibles 
	and $j$-infectious individuals. The infectious and removal rates are given in eq. 
	(\ref{eq: infection}) and (\ref{eq: removal}), which are density--dependent functions.
	
	Suppose that $Y_l$ is a independent standard Poisson process, where each 
	of possible transition $l \in \Delta$. By letting $\boldsymbol{X}_n(0)$ be non-random,
		\begin{equation}
			 \boldsymbol{X}_n(t) =\boldsymbol{X}_n(0)+ \sum_l l Y_l \left( n \int_0^t \beta_l 
			 \left( \frac{\displaystyle \boldsymbol{X}_n(s)}{\displaystyle n} \right) ds \right). 			
			 \label{eq:increment}
		\end{equation}
	Now let $\tilde{Y}_l(u) = Y_l(u) - u$ be a centered Poisson process at its expectation 
	and $\boldsymbol{\bar{X}}_n(t) = \displaystyle \frac{\boldsymbol{X}_n(t)}{n}$. 
	Following equation (\ref{eq:increment}) and letting $F(\boldsymbol{x}) = \displaystyle 
	\sum_{l \in \Delta} l \beta_l (\boldsymbol{x})$, 
		\begin{align}
			\boldsymbol{\bar{X}}_n(t) 
			&= \bar{\boldsymbol{X}}_n(0) + \frac{1}{n} \sum_{l \in \Delta} l \tilde{Y}_l \left( n 
			\int_0^t \beta_l \left( \bar{\boldsymbol{X}}_n(s) \right) ds \right) + \int_0^t 
			F(\bar{\boldsymbol{X}}_n(s)) ds. \label{eq:incrementbar}
		\end{align} 

	We will show that the stochastic process in equation (\ref{eq:incrementbar}) can be 
	approximated by non-random quantity when we let the population size become 
	sufficiently large. But before that, consider following theorem.
	
	\begin{theorem} \label{thm:fundpp}
		If $\lbrace Y(t), t \geq 0 \rbrace$ is a standard Poisson process, then for all $ s 
			\leq t$, 
		\begin{equation}
			\lim_{n \rightarrow \infty} \sup_{s \leq t} \mid n^{-1} Y(ns)-s \mid = 0 \nonumber
		\end{equation} 
		almost surely.
	\end{theorem}

	\begin{proof}
		Note that $Y(ns) \sim \text{Poisson}(ns)$, for $s \geq 0$. We can show easily show 	
		that $(Y(ns)-ns)^4$ is a submartingale. Therefore, for any $\varepsilon > 0$
		\begin{align}
			P \left( \sup_{0 \leq s \leq t} \left| Y(ns)-ns \right| \geq \varepsilon n \right) &= P \left( 
				\sup_{0 \leq s \leq t} (Y(ns)-ns)^4 \geq \varepsilon^4 n^4 \right) \nonumber \\
				&\leq \frac{E(Y(nt)-nt)^4}{\varepsilon^4 n^4} = \frac{nt (1+3nt)}{\varepsilon^4 n^2} 
				\leq \frac{C(\varepsilon, t)}{n}. \label{eq:fund.sup}
		\end{align} 
		
		If we let $n \rightarrow \infty$, then $P \left( \displaystyle \sup_{0 \leq s \leq t} \left| 
		Y(ns)-ns \right| \geq \varepsilon n \right) \rightarrow 0$. It implies that $ 
		\sum_{n = 1}^\infty P\left( \left| Y(ns) - ns \right| \right) < \infty$ since the probability of 
		its supremum tends to zero as $n \rightarrow \infty$. Therefore, according to Borel--
		Cantelli Lemma, $\displaystyle \lim_{n \rightarrow \infty} \sup_{s \leq t} \mid n^{-1} 
		Y(ns)-s \mid = 0$ almost surely.
	\end{proof}

	Now we will prove that eq. (\ref{eq:incrementbar}) can be approximated by non--random
	quantity as $n \rightarrow \infty$ by proofing theorem below.
	
	\begin{theorem} \label{thm:lln}
		Let $\displaystyle \lim_{n \rightarrow \infty} \bar{X}_n(0) = x_0$ and $x(t) = x_0+ 
		\displaystyle \int_0^t F(x(u)) du$ be the deterministic version of equation 
		(\ref{eq:incrementbar}). Suppose that for each compact $K \subset \mathbb{R}^d, 
		\exists M_K>0$ such that $\forall x,y \in K, \quad \mid F(x)-F(y) \mid \leq M_K \mid x-y 
		\mid .$ Then, for $ 0 \leq s \leq t,$
		\begin{equation}
			\lim_{n \rightarrow \infty} \sup_{s \leq t} \mid \bar{X}_n(s)-x(s) \mid = 0 \nonumber
		\end{equation}
		almost surely.
	\end{theorem}
	
	\begin{proof}
		Given $x(t) = x_0+ \displaystyle \int_0^t F(x(u)) du$, thus
		\begin{align} 
			\mid \bar{X}_n(t)-x(t) \mid 
			&\leq \mid \bar{X}_n(0)-x(0) \mid + \frac{1}{n} \sum_{l \in \Delta} \mid l \mid \sup_{s 
			\leq t} \mid \tilde{Y}_l \left( n \beta_l \left( \bar{X}_n(s) \right) \right) \mid + 
			 \int_0^t M_k \mid \bar{X}_n(s)-x(s) \mid ds. \label{eq:lln}
		\end{align}
		Now consider the second term of the right hand side of equation (\ref{eq:lln}).
		\begin{align*}
			\frac{1}{n} \sum_{l \in \Delta} \mid l \mid \cdot \sup_{s \leq t} \mid \tilde{Y}_l \left( n 
			\beta_l \left( \bar{X}_n(s) \right) \right) \mid 
			&= \sum_{l \in \Delta} \mid l \mid \cdot \sup_{s \leq t} \mid n^{-1} Y_l \left( n \beta_l 
			\left( \bar{X}_n(t) \right) \right) - \beta_l \left( \bar{X}_n(t) \right) \mid.
		\end{align*}
		According to Theorem \ref{thm:fundpp}, as $n \rightarrow \infty$, $\mid 
		n^{-1} Y_l \left( n \beta_l \left( \bar{X}_n(t) \right) \right) - \beta_l \left( \bar{X}_n(t) \right) 
		\mid \rightarrow 0$ almost surely. Therefore, equation (\ref{eq:lln}) now becomes
		\begin{align}
			\mid \bar{X}_n(t)-x(t) \mid &= \mid \bar{X}_n(0)-x(0) \mid + \int_0^t M_k \mid 
			\bar{X}_n(s)-x(s) \mid ds \nonumber \\
			&\leq  \mid \bar{X}_n(0)-x(0) \mid \cdot e^{M_K t} = 0
		\end{align}
		by Gronwall's inequallity. 
		Therefore, by sanwich property, $ \displaystyle \lim_{n \rightarrow \infty} 
		\displaystyle \sup_{s \leq t} \mid \bar{X}_n(s)-x(s) \mid = 0.$
	\end{proof}

	Theorem \ref{thm:lln} suggests that as $n \rightarrow \infty$, the process $\lbrace 
	\bar{\boldsymbol{X}}_n(t), t \geq 0 \rbrace$ will resemble the deterministic version 
	$\boldsymbol{x}(t)$. If $\lbrace Y_n(t), t \geq 0 \rbrace$ represents the number of
	infectees at time $t$ in SIS epidemic, then as $n$ sufficiently large and $t \rightarrow 
	\infty$, the process will settle down and randomly move around its mean, in this case
	are the equilibrium points of deterministic model.

\subsection{Branching Process} 

	The mimicing behaviour of stochastic model of SIS is due to the convergence of
	its process, where according to the density dependent population process, as
	$n \rightarrow \infty$ the tail of stochastic model of SIS will converge to its mean. 
	Note that after reaching some time, deterministic model of SIS will reach its
	stationary stage, i.e, extinction or endemic, which actually are the mean of
	stochastic SIS model.
	
	According to Theorem \ref{thm:lln}, all stochastic processes of SIS should be 
	mimicing the deterministic model. Obviously if $R_0 \leq 1$, both stochastic
	and deterministic models go to extinction. Unfortunately, when $R_0$ exceeds 1,
	there are some cases when stochastic model extincts. These are illustrated in 
	Figure \ref{fig:sis.(n,m)=(3e3,2)} and \ref{fig:sis.(n,m)=(3e3,10)}. In stochastic
	processes, this anomaly is not surpraising at all since given a small number of
	infectees in an early epidemic, there is non--zero chance that the individual
	will recover before it makes contact to any susceptibles. The best analogy to
	illustrate this scenario is by branching process theorem. 

Suppose that in a closed population, just by the end of its lifetime, each individual has independently produced $j \geq 0$ offsprings with probability $P_j$. Let $X_n$ denote the size of $n$-th generation, then the process $\lbrace X_n: n=0,1,2, \cdots \rbrace$ is a Markov chain having non-negative integer state space, where state $0$ is a recurrent state and others are transient. Since any finite transient state will be visited finitely often, this leads to the conclusion that if $P_j>0$, then the population will either die out or take off.

Let $\mu$ be the mean number of offspring from a single individual, then
\begin{align*}
E(X_n) &= E \left( E \left( \sum_{j=1}^{X_{n-1}} \zeta_j  \mid X_{n-1} \right)  \right), \quad \zeta_j \text{ is number of $j$th individual's offspring} \\
&= \mu^2 E(X_{n-2}) = \cdots = \mu^{n} E(X_0) = \mu^n, \quad \text{given $X_0=1$}.
\end{align*}

Suppose that $\pi_0 = \displaystyle \lim_{n \to \infty} P(X_n=0|X_0=1)$ denotes the probability that the population dies out. Note that if $\mu<1$, then $P(X_n \geq 1) \rightarrow 0$ almost surely as $n \rightarrow \infty$. This concept is very useful in epidemic modelling because when the expected number of infection is less than $1$, then the epidemic will obviously die out. But, if $\mu>1$, by conditioning on the $X_1$, $\pi_0 = \sum_{j=0}^ \infty P(\text{population dies out}|X_1=j) P_j. \nonumber$

Note that given $X_1=j$, the population will eventually die out if and only if each of the $j$ families started by the members of the first generation die out. This statement is supported by the fact that the probability of extinction occuring in the early stage is non-zero. Moreover, the probability that the members of a typical family to die out is $\pi_0$. Thus, assuming that each individual acts independently, $P( \text{population dies out}|X_1=j) =  \pi_0^j. \nonumber$

Hence, $\pi_0$ must satisfy 
\begin{equation}
\pi_0 = \sum_{j=0}^ \infty \pi_0^j P_j = E \left( \pi_0^{X_1} \right) = f(\pi_0) \label{eq:extinction}
\end{equation}
where $f(\pi_0)$ is the pgf (probability generating function) of $X_1$, in such a way that $\pi_0$ is the smallest root in $[0,1]$ satisfying equation (\ref{eq:extinction}).

\begin{table} [h]
\caption{Comparison table of probability the epidemics to die out approximated by branching process and empirical with $m=1$} \label{tab:sir.compare}
\begin{center}
\begin{tabular}{|| c c c c ||} 
\hline \hline
$R_0$		&		Population Size		&		Branching Process		&		Empiral (10000 runs) \\
\hline \hline
$\leq 1$	&		$N=100$	&	$1$		&		$1$ \\
 		& 		$N=1000$	& 	$1$		&		$1$ \\
3.0	&		$N=100$	&	$0.3333$		&		$0.3420$ \\
 		& 		$N=1000$	& 	$0.3333$		&		$0.3331$ \\
5.0	&		$N=100$	&	$0.2$		&		$0.1997$ \\
 		& 		$N=1000$	& 	$0.2$		&		$0.1996$ \\
8.0	&		$N=100$	&	$0.125$		&		$0.1273$ \\
 		& 		$N=1000$	& 	$0.125$		&		$0.1282$ \\
\hline \hline
\end{tabular}
\end{center}
\end{table}

Now suppose that $T_i$ is infection period of $i$-th typical infective, which is assumed to follow independent exponential distributions with intensity $\gamma$ and let $R$ be the total number of contacts made by a typical infective in the epidemic model. Then,
\begin{subequations}
\begin{align}
R \mid T=t &\sim \text{Poisson}(\beta t) \\
T &\sim \text{exponential} (\gamma).
\end{align}
\end{subequations}

Now, suppose that $f(s), s \in [0,1]$ is the pgf of $R$ and $g(t) \quad (t \geq 0)$ is the pdf of $T$. Then,
\begin{align*}
f(s) 
&= \sum_{k=0}^\infty s^k \int_0^\infty P(R=k \mid T=t) g(t) dt 
= \frac{\gamma}{\beta + \gamma} \sum_{k=0}^\infty \left( \frac{s \beta}{\beta + \gamma} \right) ^k.
\end{align*}
We find $s$, such that $f(s) = s$. Therefore, $s=1 \quad \text{or} \quad s = \frac{\gamma}{\beta}.$ If $\gamma \geq \beta$, then the smallest root is $1$, but if $\gamma < \beta$, then the smallest root is $\frac{\displaystyle \gamma}{\displaystyle \beta}$. Hence, suppose there are $m$ initial infectives in the population and assuming that the contacts made are mutually independent, then 
$$P ( \text{epidemic dies out} ) \approx
\begin{cases}
1 & \text{if } \gamma \geq \beta \\
\left( \frac{\displaystyle \gamma}{\displaystyle \beta} \right) ^m & \text{if } \gamma < \beta.
\end{cases} $$

Once again, the above result explains threshold behaviour of the process. All figures confirm the interpretation of threshold behaviour of the SIS deterministic model, that if $R_0 > 1$, major epidemic will occur despite of its initial values and population size. But in the stochastic model (which are displayed in jagged plots), some plots do not mimic the deterministic paths. Another simulation is provided in Table \ref{tab:sir.compare}, where we compare the probability of the epidemics to die out approximated by brancing process and empirical.

\section{Conclusion} \label{sec: conclusion}

We have successfully showed that both deterministic and stochastic models performed similar results when $R_0 \leq 1$. That is, the disease-free stage in the epidemic. But when $R_0 > 1$, the deterministic and stochastic approaches had different interpretations. In the deterministic models, both the SIS and SIR models showed an outbreak of the disease and after some time $t$, the disease persisted and reached endemic-equilibrium stage. The stochastic models, on the other hands, had different interpretations. If we let the population size be sufficiently large, the epidemic might die out or survive. There were essentially two stages to this model. First, the infection might die out in the first cycle. If it did, then it would happen very quickly, just like the branching process theory described. Second, if it survived the first cycle, the outbreak was likely to occur, but after some time $t$, it would reach equilibrium just like the deterministic version. In fact, the stochastic models would mimic the deterministic's paths and be scattered randomly around their equilibrium point.

\end{document}